\documentclass[12pt]{iopart}
\usepackage{graphicx} % Required for inserting images
\usepackage{bm}
\usepackage{svg}

\DeclareUnicodeCharacter{2009}{\,} 
\usepackage{newunicodechar}
\newunicodechar{−}{-} 
\usepackage[utf8]{inputenc}
\DeclareUnicodeCharacter{03D5}{\ensuremath{\phi}}  % 03D5 is the Unicode code for phi

\usepackage{tikz} %tikz is used to draw the glyphs
\usetikzlibrary{arrows,scopes} %tikz libraries used to draw the glyphs
\begin{document}
\title[]{Improved Bethe-Heitler positron creation and retention by combining direct laser acceleration and solid target interaction within a gas jet}
\author{Lucas Ivan I\~nigo Gamiz$^1$, Robert Babjak$^{1,2}$, Bertrand Martinez$^{1}$, Marija Vranić$^1$}

\address{$^1$ Golp/Instituto de Plasma e Fusão Nuclear, Instituto Superior Técnico, Universidade de Lisboa, 1049-001 Lisbon, Portugal}
\address{$^2$ Institute of Plasma Physics, Czech Academy of Sciences, Za Slovankou 1782/3, 182 00 Praha 8, Czechia}
\ead{lucas.inigo.gamiz@tecnico.ulisboa.pt}

\begin{abstract}
    The next generation of Petawatt-class lasers presents the opportunity to study positron production and acceleration experimentally, in an all-optical setting. Several configurations were proposed to produce and accelerate positrons in a single laser stage. However, these configurations have yielded limited positron beam quality and low particle count. This paper presents methods for improving the injection and retention of positrons obtained via Bethe-Heitler pair production and accelerated using direct laser acceleration (DLA) in a plasma channel. The work first introduces a semi-analytical model which predicts laser energy depletion in this highly nonlinear regime. We demonstrate through PIC simulations that accelerated electrons can induce charge inversion within the channel, leading to positron trapping and acceleration. We investigate how laser focusing position, channel wall density, target foil position and target thickness influence positron creation and retention. Our configuration can achieve an 8-fold increase in positron retention compared to previous studies and a higher number of positrons produced overall. This work establishes a robust, single-stage approach for obtaining positron beams, opening new avenues for experiments with Petawatt-class lasers and potential applications in electron-positron collisions and QED cascades.
\end{abstract}

\maketitle

\section{Introduction}

The acceleration of particles has a wide variety of applications in high-energy physics \cite{bulanov_electromagnetic_2013,greaves_electron-positron_1995}, astrophysics \cite{wardle_electronpositron_1998,mirabel_sources_1999} and for radiation sources \cite{benedetti_whitepaper_2024}. Traditional radio-frequency (RF) accelerators have successfully accelerated particles to high energies. Electron-positron pairs were created experimentally by accelerating electrons in a linear accelerator to 46.6 GeVs and colliding them against a laser \cite{burke_positron_1997}. This seminal experiment successfully demonstrated that a laser and electron beam interaction can create positrons, but the positron yield was relatively low, around $100$ positrons for the entire campaign. Further experiments are needed to produce electron-positrons pairs in the strong field regime, where the number of created particles would be higher by several orders of magnitude and one could potentially study the collective effects of lepton plasmas. Due to the size and costs of radio-frequency accelerators, there is a need for smaller and more accessible machines to perform these experiments
\\
\\
To address these limitations, researchers have turned to plasma-based accelerator techniques due to their tolerance to high acceleration gradients. Two prominent examples are laser-wakefield acceleration (LWFA) and direct laser acceleration (DLA). LWFA uses intense short laser pulses to generate plasma waves, which can accelerate electrons to high energies at a rate of GV/m, shrinking the acceleration distance from kilometers, to centimeters \cite{tajima_laser_1979}. LWFA has demonstrated to accelerate quasi-monoenergetic electron beams to high energies and with low divergence \cite{mangles_monoenergetic_2004,wang_petawatt-laser-driven_2012,kim_enhancement_2013,gonsalves_petawatt_2019,aniculaesei_acceleration_2023}. 

On the other hand, DLA acceleration is a resonant process due to the coupled oscillations in the fields of a laser pulse and betatron oscillations in the background ion channel fields created in the plasma by electron displacement \cite{pukhov_particle_1999, pukhov_relativistic_1998, vranic_extremely_2018}. The main advantage of this scheme is that it allows accelerating electron bunches with charge over 100 nC \cite{hussein_towards_2021, shaw_microcoulomb_2021} over a few millimetres of laser propagation, which is 1000 times higher charge compared to LWFA sources. Recently, it has been demonstrated that optimal focusing strategy enables DLA acceleration of electrons up to GeV energies with conversion efficiency in tens of percent \cite{babjak_direct_2024}. This has also been proved experimentally \cite{tang_influence_2024}. The drawback of the scheme is a broad Maxwell-like spectrum and higher divergence in tens of millirad. However, the high conversion efficiency and high charge make the mechanism promising for applications which do not require monoenergetic electrons such as bright X-ray sources \cite{kneip_2008, cipiccia_gamma-rays_2011,cikhart_2024,rinderknecht_2021}, neutron \cite{gunther_2022, cohen_undepleted_2024} and positron generation \cite{warwick_experimental_2017,sarri_generation_2015}  or seeding pair cascades \cite{bell_2008, nerush_2011, grismayer_2017}.
\\
\\
The imminent commission of next-generation Petawatt laser facilities has sparked significant interest in positron creation and acceleration \cite{yoon_achieving_2019,papadopoulos_first_2019,bromage_technology_2019}. With these laser intensities, we can create conditions in which electron-positron pair production can occur due to the scattering between a laser and an ultra-relativistic electron beam \cite{sokolov_pair_2010,lobet_generation_2017,blackburn_scaling_2017,streeter_narrow_2024} or the laser interacting with a solid target \cite{krajewska_bethe-heitler_2013}. Theoretical studies have demonstrated the production of positrons by the interaction between an ultra-relativistic electron beam and Petawatt lasers \cite{sokolov_pair_2010,blackburn_scaling_2017} as well as a single-stage production and acceleration of positrons in vacuum \cite{vranic_multi-gev_2018}. Besides accelerating electrons, wakefield accelerators have been proposed to create \cite{gahn_generation_2002,sarri_generation_2015,sarri_table-top_2013} and accelerate positrons \cite{vieira_nonlinear_2014,yi_positron_2014,diederichs_positron_2019,silva_stable_2021,silva_positron_2023}. Several experiments were performed demonstrating the feasibility of positron acceleration in WFAs \cite{muggli_halo_2008,streeter_narrow_2024}. However, while LWFA is effective for electron acceleration, its guiding structure tends to defocus positrons \cite{schroeder_physics_2010,joshi_perspectives_2020}, necessitating alternative approaches that can efficiently accelerate and retain positively charged particles.
\\
\\
Our team has previously proposed single-stage schemes based on DLA to create and accelerate positrons obtained via Breit-Wheeler \cite{martinez_creation_2023,maslarova_radiation-dominated_2023} and Bethe-Heitler \cite{martinez_direct_2024} pair production. In contrast to LWFA, the acceleration mechanism from DLA is agnostic to the particle charge. Furthermore, it can potentially retain more positrons by creating a focusing structure for positively charged particles. A previous study showed that DLA can accelerate positrons to GeV energies, but the accelerated positron beam loses approximately 99.8$\%$ of charge within a millimetre of propagation from the creation point \cite{martinez_direct_2024}. Optimization studies are needed to improve positron beam retention and make DLA a viable candidate for acceleration.
\\
\\
In this work, we present a strategy to improve a scheme for single-stage positron production, acceleration and retention of the positron beam using next-generation Petawatt lasers and DLA. This new strategy allows for more than an order of magnitude greater number of detectable positrons (created and retained). A 10 PW laser, with the parameters available at ELI-beamlines facility (L4 laser), is directed into a plasma channel, propagates through it, and self-guides. As it propagates, electrons are collected by the laser and accelerated by DLA. Later, the laser and the electrons collide with an aluminium foil perpendicular to the laser propagation direction, producing photons by bremsstrahlung and producing pairs via Bethe-Heitler. This is similar to when LWFA electrons are accelerated and collide with a target \cite{alejo_laser-wakefield_2019}. However, using DLA, we can accelerate more electrons, potentially creating more pairs. We can also guide the positrons and even accelerate them after creation. We demonstrate that the number of created and retained positrons depends on the laser focusing position, plasma density, foil position and thickness. By placing the foil at a position where the laser ability to accelerate charge saturates, we maximize the number of pairs created and improve the focusing structure for the positrons.
Consequently, this structure will retain over 10 per cent of the positrons after one millimetre of laser propagation. A small fraction of positrons can accelerate up to GeV energies, which can potentially be improved by applying a similar optimization strategy as for electrons \cite{babjak_direct_2024}.
\\

\begin{figure}[ht]
    \centering
    \includegraphics[width = 0.8\textwidth,height = 0.4\textwidth]{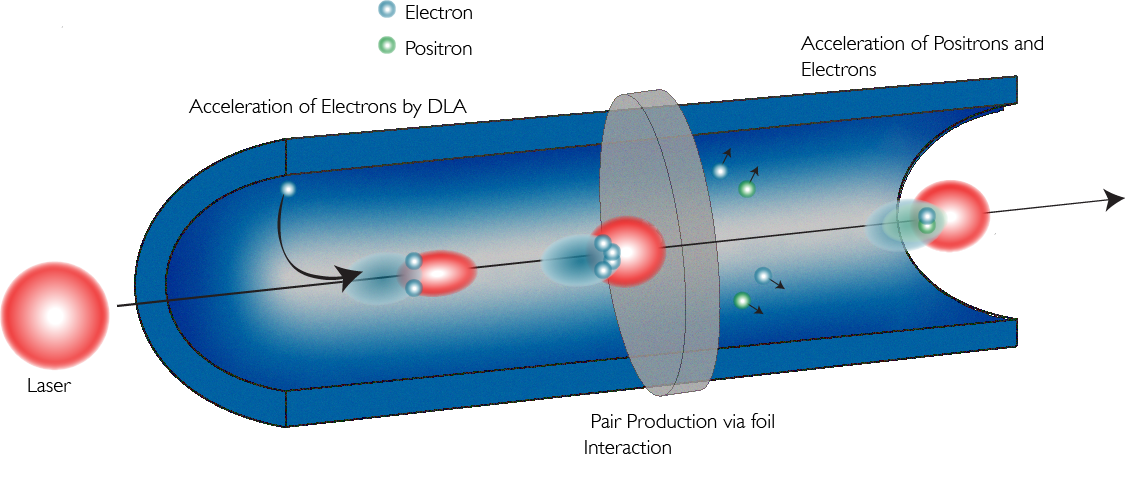}
    \caption{Setup depicting the stages of the simulation. i) A laser beam entering the plasma channel. ii) Laser propagating through the channel, capturing some electrons from the channel wall and accelerating them via direct laser acceleration. iii) Collision between the laser/electron beam and the foil. iv) guiding and acceleration of electrons and positrons after the collision.}
    \label{setup}
\end{figure}

The proposed configuration is illustrated in Fig.~\ref{setup}. As the laser enters the plasma channel, electrons within the channel displace due to the ponderomotive force exerted by the laser. As the laser propagates through the channel, it relativistically self-focuses and self-guides. A fraction of the electrons from the wall are injected into the channel undergo betatron oscillations and experience DLA  \cite{babjak_direct_2024,martinez_creation_2023,maslarova_radiation-dominated_2023}. As the laser continues to propagate, more charge is injected and accelerated at the expense of the laser energy. Once the injected electron charge exceeds the background ion density, the transverse field sign reverses, creating a structure which can focus and guide positrons. Later, the electrons and the laser collide with the aluminium target, creating photons by Bremmstrahlung \cite{koch_bremsstrahlung_1959} and nonlinear Inverse Compton scattering \cite{reiss_absorption_1962}. The photons can decay into electron-positron pairs via Bethe-Heitler \cite{motz_pair_1969} or Breit-Wheeler \cite{breit_collision_1934} processes. Most of these newly produced particles propagate along the laser propagation direction. As the DLA accelerated charge passes through the foil and produces photons, consecutively producing pairs, the focusing structure created by the accelerated electrons retains a fraction of positrons. The retained positrons gain energy from the laser and perform resonant motion, resulting in DLA just like electrons.

This paper is structured as follows: Section 2 summarizes quasi-3D simulation parameters for the simulations performed using OSIRIS \cite{fonseca_osiris_2002}. Section 3 introduces DLA and demonstrates the creation of a guiding structure using quasi-3D PIC simulations. In Section 4, we introduce a semi-analytical model, obtained using the least square regression technique for the energy transfer of the laser to the electrons and the background plasma. In section 5, we investigate the effect of the laser focusing position inside the channel on the number of positrons created and retained. Later, we fix the position of the laser focal spot and then vary the position of the foil, highlighting the effects of the beam loading and on positron production. Then, we change the number density and place the foil further into the channel to saturate the amount of charge the laser can accelerate. Finally, we vary the foil thickness. Lastly, We present the conclusions and the best viable setup to produce and retain the most positrons for different applications and scenarios.

\section{Quasi-3D simulations}

Simulating the proposed setup in a complete 3D simulation is unfeasible due to the disparity of the spatial and temporal scales. Alternatively, quasi-3D simulations can be used by exploiting the cylindrical symmetry of the system, greatly reducing the computational resources required. The advantage of Quasi-3D over 2D geometry is that it can describe the correct 3D laser dynamics and allow for quantitative predictions of different observables (e.g., charge). We use the quasi-3D module of the particle-in-cell code \verb|OSIRIS| \cite{fonseca_osiris_2002,davidson_implementation_2015}. The simulations also contain QED modules that calculate Bremsstrahlung \cite{koch_bremsstrahlung_1959}, Breit-Wheeler \cite{breit_collision_1934}, Bethe-Heitler \cite{motz_pair_1969} and Nonlinear Inverse Compton Scattering \cite{reiss_absorption_1962}. In the simulations, all these processes are activated. 
For each simulation, the plasma is initialized as a preformed channel with a number density profile $n(r) = n_p + \left(n_w - n_p \right) \left( r / r_c \right)^\alpha$, where $n_p$ is the plasma density at the centre of the channel, $n_w$ is the density at the wall of the channel, $r$ is the radial coordinate, $r_c$ is the radius of the plasma channel, and $\alpha$ is an integer denoting the steepness of the plasma profile, which in here is set to $5$. In each corresponding section, we will specify the wall and background densities. The laser has a Gaussian profile with a peak intensity of $5\times10^{24}~$W/cm$^{-2}$, $1~\mu$m wavelength, a duration of $150~$fs, and a waist of $3.4~\mu m$. The aluminium foil is set as pre-ionized, with a density of $3\times 10^{22}$~cm$^{-3}$. We assume that the foil is pre-expanded, where we have an exponential pre-plasma layer with a length of 500 nm, added to the front of the foil.
The simulation domain is $137.5 \times 80~\mu$m$^2$ cell size is $dr = dx = 16~$nm, with one angular mode and 26.6 attosecond timestep. 16 particles per cell represent the electrons and ions in the channel, and 32 particles per cell represent the electrons and ions in the aluminium foil. The simulation has a moving window co-propagating with the laser.

\section{Direct Laser Acceleration and creation of a guiding structure}

In DLA, electrons are accelerated within a background channel. The ions produce a radial electric field created by charge separation and an azimuthal magnetic field produced by the accelerated charge \cite{vranic_extremely_2018, jirka_scaling_2020, pukhov_particle_1999}. In the vicinity of the central axis, the channel fields linearly depend on the radial distance from the centre. The transverse force attracts electrons which oscillate around the channel axis with a betatron frequency $\omega_{\beta}=\omega_p/\sqrt{2\gamma}$. Resonant matching of betatron frequency and the Doppler-shifted laser frequency can result in acceleration to energies of several GeV with multi-PW facilities \cite{babjak_direct_2024}. The continuous injection of electrons from the channel wall into the field of a laser pulse \cite{valenta_direct_2024} can result in electron beam loading that inverses the background field polarity, making a structure which can confine positrons \cite{martinez_creation_2023, martinez_direct_2024}. 
It has been shown previously that the maximum attainable energy in the DLA regime depends on the oscillation amplitude \cite{babjak_direct_2024,babjak_direct_2024-1}. This amplitude is defined by the integral of motion $I = \gamma - p_x/(m_ec) + \omega_p^2 y^2 / 4c^2$ which is a conserved quantity during laser-particle interaction if no radiative losses or density variations are present \cite{arefiev_parametric_2012, khudik_universal_2016, babjak_direct_2024-1}. The resonant condition $\epsilon_{cr} < a_0 \omega_p / (\omega_0 I^{3/2})$ \cite{khudik_universal_2016,arefiev_enhancement_2014} limits the transverse oscillation amplitude capable of being resonantly accelerated \cite{babjak_direct_2024}, thereby limiting the energy gain. Since electrons with higher oscillation amplitude can reach higher energies, the most energetic particles are those with the highest resonant oscillation amplitude.

\begin{figure}[ht]
\centering
\includegraphics[keepaspectratio, width = 0.9\textwidth]{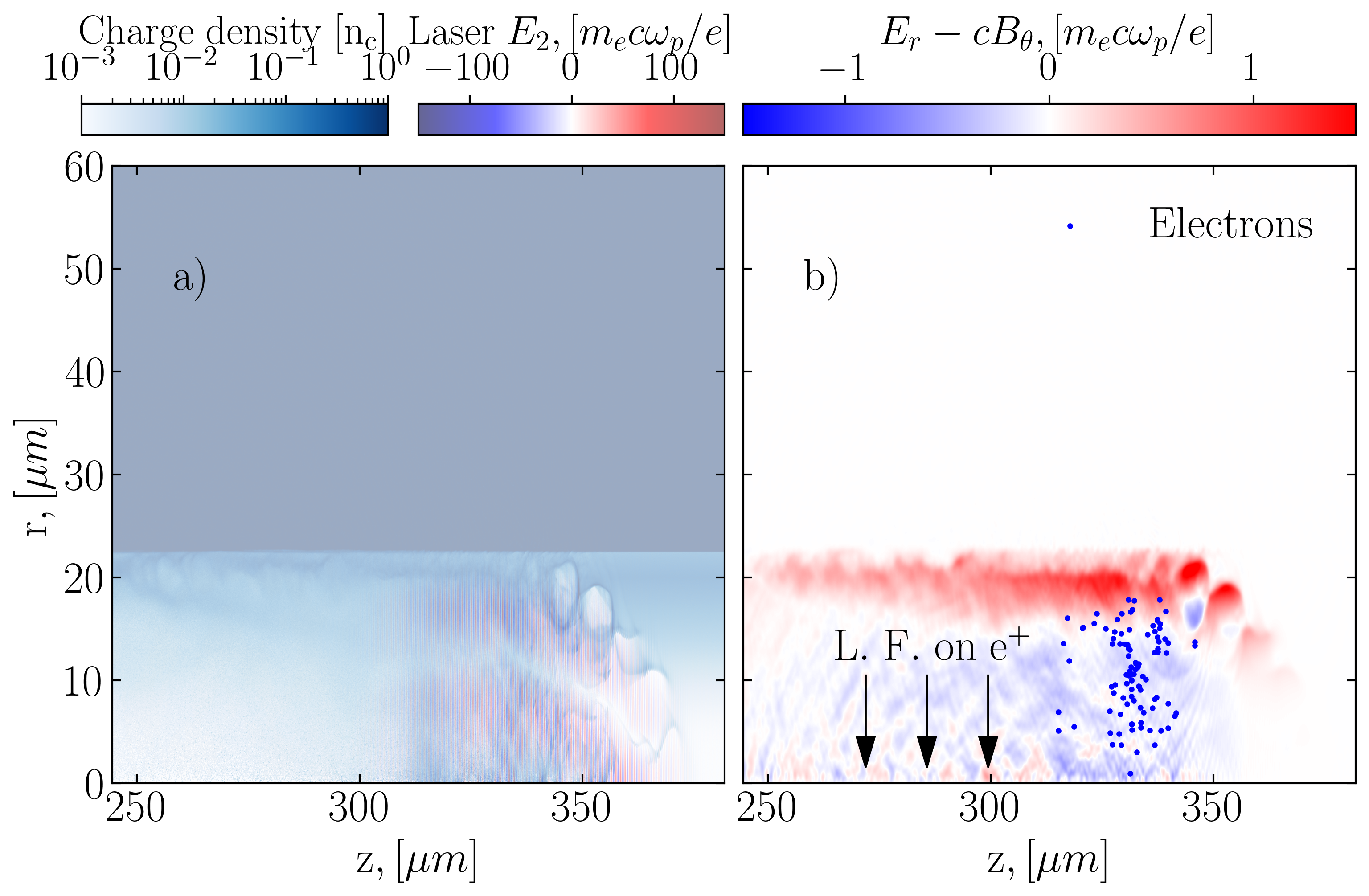} 
 \caption{A close-up of a), the electron density of the plasma channel and the laser at an early stage of the simulation. b) The field inversion corresponds to an accelerated electron beam and 100 randomly sampled electrons as blue dots. A black arrow is added to depict the force the positrons will experience.}
\label{fieldInversion}
\end{figure}

Fig.~\ref{fieldInversion} illustrates a snapshot of a simulation showing electron beam loading after the laser has propagated 375 micrometres through the plasma channel. In panel a), we see the laser field superimposed with the electron density of the plasma channel. The laser is guided within the channel, and one can observe how the laser displaces electrons and how the charge increases towards the channel axis. Frame b) showcases the formation of the guiding structure, coloured in blue, highlighting the positions of some accelerated electrons within the channel. 
As the laser accelerates more electrons, the fields generated by this structure can surpass those of the ion background. In this case, the resulting Lorentz force acting on positrons is directed towards the channel axis. In other words, the electron injection leads to field inversion, where the transverse electric fields now act to focus positrons. However, positron confinement is not absolute. Positrons possessing energy exceeding the limit that allows crossing the potential barrier created by these transverse fields can escape. 

\section{Description of energy transfer from the laser to the electrons}

As the laser propagates through the plasma channel, it exerts a ponderomotive force on plasma electrons, creating an ion cavity. Later, the electrons are pulled back and can be attracted by the radial electric field of the ion channel. The laser can then accelerate some of these electrons. However, the laser capacity to accelerate charge is finite \cite{babjak_direct_2024}. The electron injection and beam loading were, explored previously in detail \cite{martinez_direct_2024, valenta_direct_2024}. 
Here, we aim to account for the total energy transferred from the laser to the background plasma and the accelerated electrons. Based on earlier works, this is not a linear process and its full description may be out of reach. As the laser interacts with and accelerates electrons, it transfers energy to the electron beam. The laser beam loses more energy with an increasing number of accelerated particles. But this accounts only for a small fraction of the laser energy loss (typically a few percent). The laser also invests energy to create a plasma channel. Additional dissipation can occur due to streaming and surface instabilities. When the laser is depleted, there can be no further acceleration.
\\
Investigating the rate at which the laser transfers energy to the plasma is not trivial, as many factors can contribute to it. However, we can paint a simplified picture of the evolution of laser energy. As the laser propagates through the plasma, the forefront of the laser will begin to deplete, being the first to interact with the plasma. In this model, we assume uniform plasma, that the rate of energy transfer per unit of time is a function of laser intensity $a_0$, plasma density, $n_p$, and laser duration $\tau$. In our case, we assume that these parameters are slowly varying. 
We take a laser to occupy a cylindrical volume with transverse spotsize $S$, and length $c \tau$. The energy contained in a plasma within the laser volume is then
\[
\mathcal{E}_l = a_0^2 S c\tau n_c
\]
where $n_c$ is the critical density. Only a fraction of the laser interacts with fresh plasma in an infinitesimal time, $dt$. The volume corresponding to this fraction is $scdt$, which contains the energy $a_0^2Scn_pdt$. A part of this energy can be absorbed. Therefore, we expect
\[
d\mathcal{E}_l\propto \mathcal{E}_l \frac{dt}\tau
\]
where the right hand side denotes the available energy. The percentage of the energy absorbed is expected to depend on specific values of $a_0, n_p$ and $\tau$. One can define a coefficient $k$ such that 
\begin{equation}
d \mathcal{E}_l \propto -k\mathcal{E}_l\frac{dt}{\tau}.\label{laserEnergyRate}
\end{equation}
where $k = f(a_0, n_p)$ note that $k$ is constant as long as the laser and plasma parameters do not significantly change over time if we consider constant plasma density. 
 The conservation of energy equation of the system is given by
\begin{equation}
\mathcal{E}_\mathrm{total} = \mathcal{E}_l(t) + \mathcal{E}_e(t) + \mathcal{E}_\mathrm{loss}(t).
\end{equation}
Where $\mathcal{E}_\mathrm{total}$ is the total energy of the system, $\mathcal{E}_l(t)$ is the laser energy as a function of time, $\mathcal{E}_e(t)$ is the electron beam energy as a function of time and $\mathcal{E}_\mathrm{loss}(t)$ is the energy ``lost" to the creation of the plasma channel. Since the laser is the only source of energy in the system, we can describe the initial conditions as:
\begin{equation*}
\mathcal{E}_\mathrm{total} = \mathcal{E}_l(0) = \mathcal{E}_{l,0}, \hspace{2cm} \mathcal{E}_e(t) = \mathcal{E}_\mathrm{loss}(t) = 0.
\end{equation*}
After a long time, most of the laser energy will be transferred to the electrons in the plasma or the beam or lost $\mathcal{E}_l(t \rightarrow \infty) \approx 0$. After a long time, when the laser is fully depleted, our conservation of energy equation takes the following form:
\begin{equation}
\mathcal{E}_{l,0} = \mathcal{E}_e(t \rightarrow \infty) + \mathcal{E}_\mathrm{loss}(t \rightarrow \infty)
\end{equation}
The solution to the differential equation describing the laser depletion (Eq.~\ref{laser_energy_eqn}) with the initial conditions described yields
\begin{equation}
    \mathcal{E}_l = \mathcal{E}_{l,0}e^{-kt/\tau} \label{laser_energy_eqn}
\end{equation}
This solution is valid as long as $a_0, \tau$ and $n_p$ do not change significantly. One can expect this to be a reasonably good assumption until the laser has lost about $50\%$ of its energy. We, therefore, assume that $k$ depends only on the initial $a_0$ and the laser duration $\tau$ and that the laser propagates through constant density plasma. This way, we can hope to obtain a phenomenological fit for a function describing $k$ using multiple particle-in-cell simulations in the parameters of interest for DLA with next-generation lasers.
We perform numerous simulations varying the laser power, $1, 5, 10$ PW, plasma densities ($0.3, 0.2, 0.1, 0.01~n_c$), with the laser waist of $8.0~\mu$m and the laser duration 200 fs. We then use the Python library \verb|SciPy| \cite{virtanen_scipy_2020} and the method \verb|curve_fit|, which uses non-linear least squares regression, to fit the laser energy and the average energy of the electrons over time to the energy transfer rate $k$. To find the dependence of the rate $k$ to the number density and the laser strength; we assume a polynomial function with the following shape:
\begin{equation}
k_\mathrm{fit} = A (n_p/n_c)^a a_0^b
\end{equation}
Where $A$ can be a function of the laser spotsize, and the laser envelope profile. Since these are constant in our simulations, $A$ is assumed constant here. We find the fitting parameters $A = 2.06 \pm 0.71, a = 0.48 \pm 0.06, b = -0.50 \pm 0.09$. The opposite sign suggests that $k\approx A\left(\frac{n_p}{a_0 n_c}\right)^a$ which makes physical sense as the ratio $n_p / (a_0 n_c)$ represents the ratio of plasma density to the relativistically corrected critical density. This ratio has been previously introduced in the literature as a ``self-similarity parameter" which is useful for describing laser-plasma interactions and predicting electron dynamics in the presence of very intense laser fields \cite{gordienko_scalings_2005}. We have not assumed self-similarity here, but it did emerge naturally as a result of our analysis, The exponent ``$a$"  for total absorption is not known, and our fit suggests it is ~$0.5$. This simple model illustrates how the laser energy transfer depends on the plasma density and the laser strength. A more intense laser can take more time to transfer its energy to a plasma as it provides more energy. Similarly, if the plasma is more dense, it can deplete the laser energy faster.

\begin{figure}[ht]
\centering
\includegraphics[keepaspectratio, width = 0.9\textwidth]{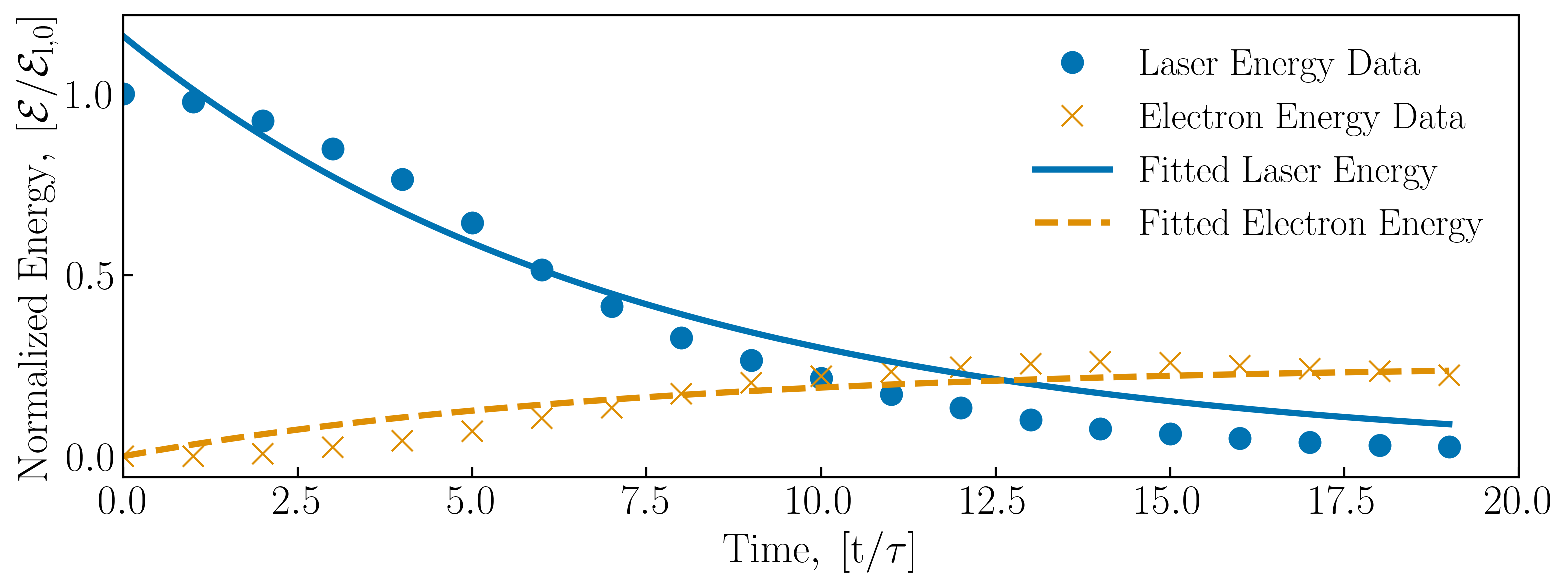}
\caption{The laser energy evolution plotted alongside the electron energy data and a fitting of the equation \ref{laser_energy_eqn} with data from a simulation with parameters: laser power = $1$ PW, $a_0 = 27, n_p = 0.1~n_c$, laser duration $\tau = 200~$fs and laser waist $8.0~\mu$m. The entrance to the channel has a 10~$\mu$m long gradient.}
\label{sim_fit}
\end{figure}

 This equation helps engineer an ideal setup, and generate and accelerate Bethe-Heitler positrons as it gives an estimate on the degree of the laser depletion over time and the propagation distance.

\section{Results}
\subsection{The effect of the laser focal position on the number of created positrons}

Martinez \textit{et al.} \cite{martinez_direct_2024} showed that a laser with an $a_0 = 200$ and 150~fs duration loads enough electron charge into the plasma channel approximately at $200~\mu$m of laser propagation. Therefore, we fix the aluminium foil at $200~\mu$m inside the channel to maximize the electron loading and with that, enhance the production of positrons. Furthermore, we vary the laser focal position inside the plasma channel to investigate the laser focusing effects on the creation and retention of positrons. In these simulations, the wall density $n_w = 20~n_c$ and the background density $n_p = 0.001~n_c$. In reality, creating such a density profile is challenging, in here, we provide a comprehensive overview of the proposed scheme to accelerate positrons. Using Eq.~10, we can obtain the approximate laser energy when it hits the foil $200~\mu$m inside the plasma channel to be $96\%$ of its original strength. This value is expected to depend on the laser guiding. 

\begin{figure}[ht]
    \centering
    \includegraphics[keepaspectratio,width = 0.9\textwidth]{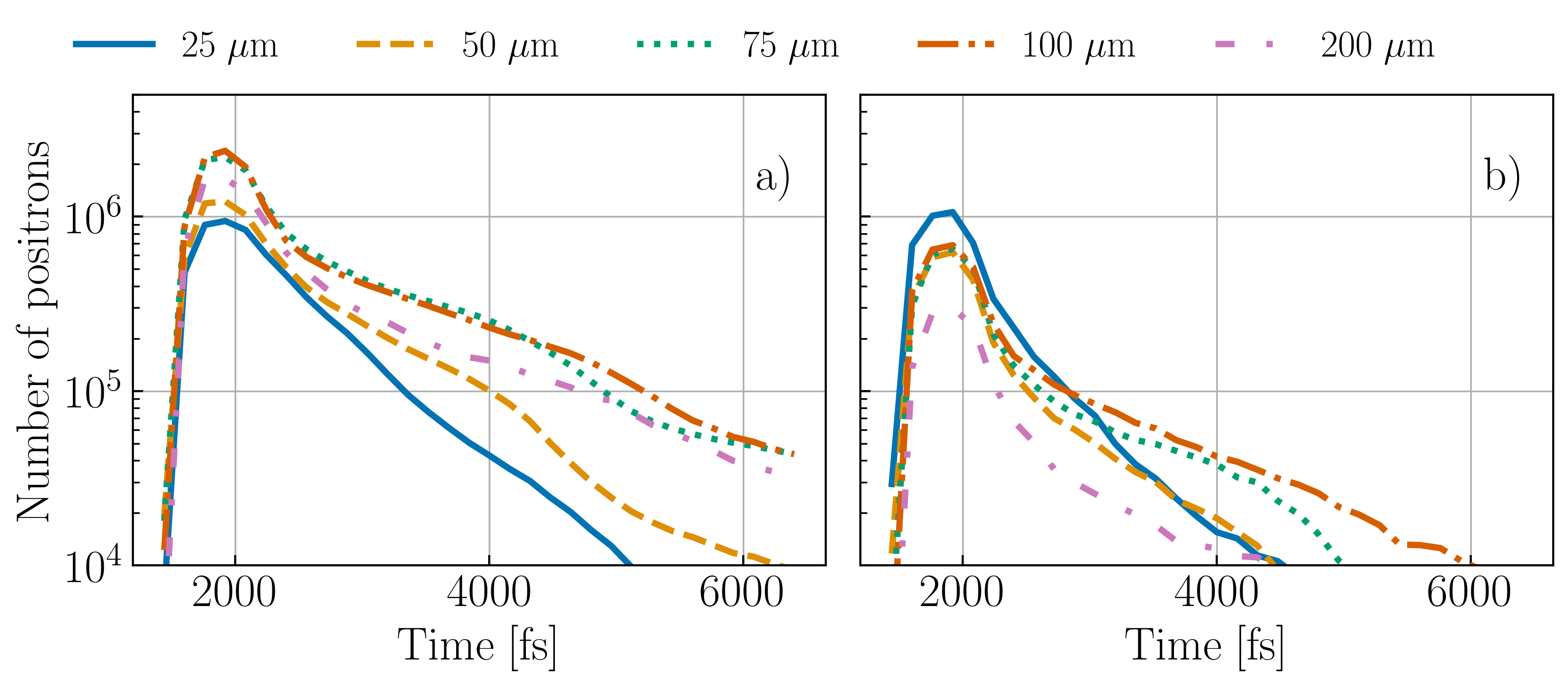}
    \caption{Number of created positrons by fixing an aluminium target $200~\mu$m inside the plasma channel and changing the focusing position of the laser at $25, 50, 75, 100, 200~\mu$m inside the channel. Panel a) shows the number of positrons created by the accelerated electrons. Panel b) shows the number of positrons the laser creates interacting directly with the aluminium foil (foil electrons emitting photons within the target)}
    \label{num_acc_pos_comp_fp200}
\end{figure}

Figure~\ref{num_acc_pos_comp_fp200} compares the total amount of accelerated positrons created from the channel and the foil particles. It demonstrates that the most significant contribution to the accelerated positrons arises from the DLA accelerated electron charge impacting the aluminium foil, creating bremsstrahlung photons, which then produce pairs via Bethe-Heitler rather than the direct interaction between the laser and the foil, where the pairs would originate from the foil electrons. Figure~\ref{num_acc_pos_comp_fp200} also shows that the laser focusing point directly affects the amount of the electron charge, thus affecting the total positrons created. We can also infer that if the laser focusing point is close to the channel entrance, the laser will be better guided, making slightly more pairs from the laser-foil interaction than the accelerated charge. If the laser is focused around $75~\mu$m after the channel entrance, we do not compromise much of the laser energy. Still, we can accelerate enough electrons at high energies such that the production of positions substantially increases.

\begin{figure}[ht]
    \centering
    \includegraphics[keepaspectratio,width = 0.9\textwidth]{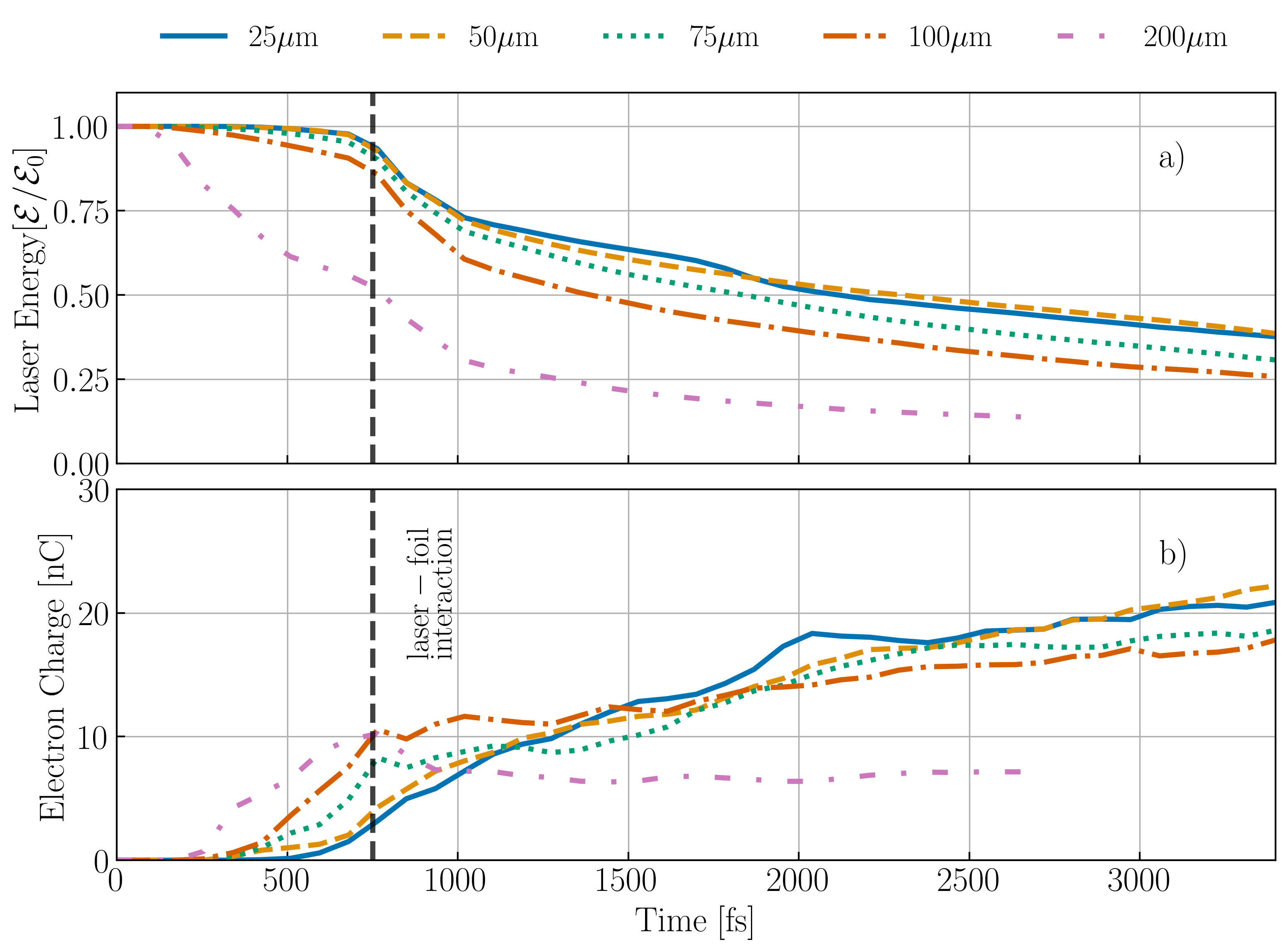}
    \caption{Simulation results for different laser focusing positions (25, 50, 75, 100, 200 $\mu$m) inside the plasma channel. Panel a) shows laser energy evolution over the simulation of each different focusing position. Panel b) shows accelerated charge evolution along the simulation for each focusing position.}
    \label{acc_charge_comp}
\end{figure}

Figure~\ref{acc_charge_comp} shows how the laser focal spot position influences the accelerated electron charge. In DLA, the laser spot evolution plays a crucial role because the most efficient acceleration is present when the spot size is matched to the maximum resonant oscillation amplitude \cite{babjak_direct_2024}. As the focal spot is set further into the channel, the laser transverse spotsize is larger as it enters the plasma channel. This permits the laser to interact with a larger volume, thus accelerating more charge (observed in panel b) from Fig~\ref{acc_charge_comp}). However, as the laser interacts with more electrons, it loses energy faster. Besides, the accelerated electrons are lost more rapidly in this case. This can also be observed in panel a), where the laser energy is lost considerably faster for the simulations where the laser focuses further into the channel. 

Focusing the laser deeper into the channel changes the effective spot size, which affects the beginning of electron acceleration. It has been previously shown \cite{babjak_direct_2024} that a wider laser spotzise increases the conversion efficiency and enhances electron acceleration if the laser waist is smaller than a maximum transverse displacement where the electrons can be accelerated efficiently, defined as $y_\mathrm{max}$. Thus, a larger spotsize results in more energetic electrons. In a low background plasma density, $y_\mathrm{max}$ exceeds 10~$\mu$m. Consequently, a larger spot size will result in more energetic electrons. A higher charge and more energetic electron beam will produce more pairs by interacting with the target. Nonetheless, the laser spotsize has to be carefully selected to optimize the laser energy transfer of the laser to the electrons \cite{babjak_direct_2024} and maximize the accelerated charge.

\subsection{The effect of foil positioning on positron generation and retention}

The aluminium foil can be placed deeper inside the plasma channel to maximise the DLA-accelerated charge. This allows the laser to accelerate a significant charge to high energies to maximize the positrons created by the charge-foil impact. The further the foil is from the channel entrance, the more particles the laser accelerates before the interaction with the target.

We can place the foil further into the channel, aiming to maximise the laser capabilities of acceleration. Babjak~\textit{et al.} mention that lower-density plasma channels can lead to higher energies of the accelerated particles \cite{babjak_direct_2024}. The number density of the wall we use here is $0.2~n_c$, and the plasma background density is $0.001~n_c$. Using these parameters, we can expect the accelerated electron energy saturation to occur at approximately 1500~MeV and at 200 microns inside the channel. However, we expect radiation losses by the accelerated electrons, effectively decreasing the acceleration gradient; Besides, we expect a contribution of the longitudinal fields to the maximum energy attained by the electrons \cite{wang_direct_2019}.
Numerous electrons create the guiding structure for positrons. Creating the positrons with pre-existing guiding fields may result in better retention of the injected charge. From Eq.~\ref{laser_energy_eqn}, and scaling the parameter $A$ with the ratio of the waist of the pervious simulations to this $3.4~\mu$m, we can expect the laser energy remaining to be $92\%, 87\%, 83 \%, 79\%$ and $75\%$ for the foil positions at $400, 600, 800, 1000,$ and $1200~\mu$m, respectively, inside the channel. We expect the laser to have depleted over $99\%$ of its energy after 10 mm of propagation.

\begin{figure}[ht]
\centering
\includegraphics[keepaspectratio,width = 0.9\textwidth]{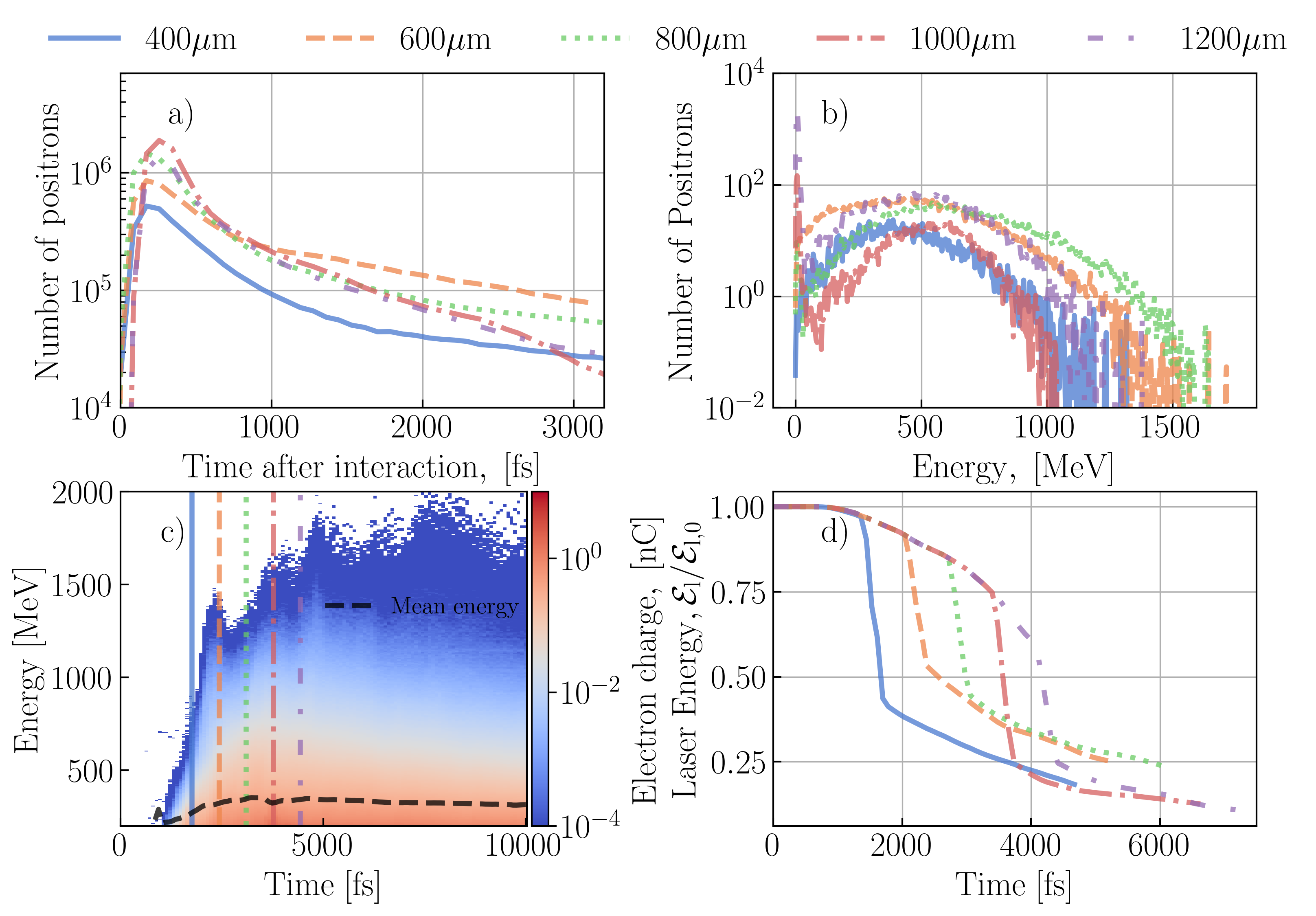}
\caption{ a) total positrons created and accelerated for a foil $400, 600, 800,$and $1200~\mu$m inside the plasma channel for a wall density of $0.2~n_c$ and a background density of $0.001~n_c$. b) energy spectra of the positrons at the last timestep of the simulations for the foil as mentioned above positions and densities. c) is a waterfall plot of the accelerated electron charge distribution with the vertical lines coloured and located to correspond to the time the laser interacts with the foil and a dashed black line showing the mean energy. d) is the laser energy over the simulation for the simulations for the foil located at $400, 600, 800, 1000$ and $1200~\mu$m.}
\label{comparisons_further_channel}
\end{figure}  

In Fig.~\ref{comparisons_further_channel}, one can confirm that there are simulations which produce more positrons than others. Besides, we can see that we retain the positrons better than in the previous section. It can also be inferred that, in this setup, there is an optimal position for the aluminium foil. When the foil is placed too early, the laser, electrons, and foil interaction do not create many positrons. On the other hand, if placed later, we can increase the number of created positrons and improve their retention. However, a target placed even further into the channel decreases the maximum number of created positrons and the percentage of retained positrons. The loss can be attributed to the laser losing energy as it propagates through the plasma. In other words, by placing a foil deep in the channel, it is possible to generate more energetic electrons that can increase the positron yield, but the laser is too depleted to further guide and accelerate the positrons. Furthermore, in panel b) in Fig. \ref{comparisons_further_channel}, it is observed that the positrons with the foil placed $800~\mu$m inside the channel are more energetic, with energies up to $1.5~$GeVs. Thus, it is needed to have sufficient laser energy and sufficient accelerated electron charge to both, retain positrons and guide the positrons after interacting with the target. Frame c) shows the electron energy distribution evolution for a simulation where there is no foil inside the channel. Using the theory developed by Babjak~\textit{et al}\cite{babjak_direct_2024} we obtain that the electron acceleration should stop at approximately 200~$\mu m$, and the electrons reach a theoretical maximum energy of $1.5~$GeV. Here, we see that the electron acceleration stops between $600$ and $800~\mu$m inside the channel. The difference between the expected saturation at 200 microns and our results can be attributed to our tightly focused regime, where this theory does not account for radiative losses. We let the simulation run an extra millimetre after the target interaction so that the positrons can have the same time to reach these theoretical limits. Frame d) shows how the laser loses energy as it propagates and how much it yields to the foil. These can be compared with the energy of the positrons shown in frame b) which are more energetic if the laser has more energy available to transfer to the positrons. 

\begin{figure}[ht]
\centering
\includegraphics[keepaspectratio,width = 0.9\textwidth,]{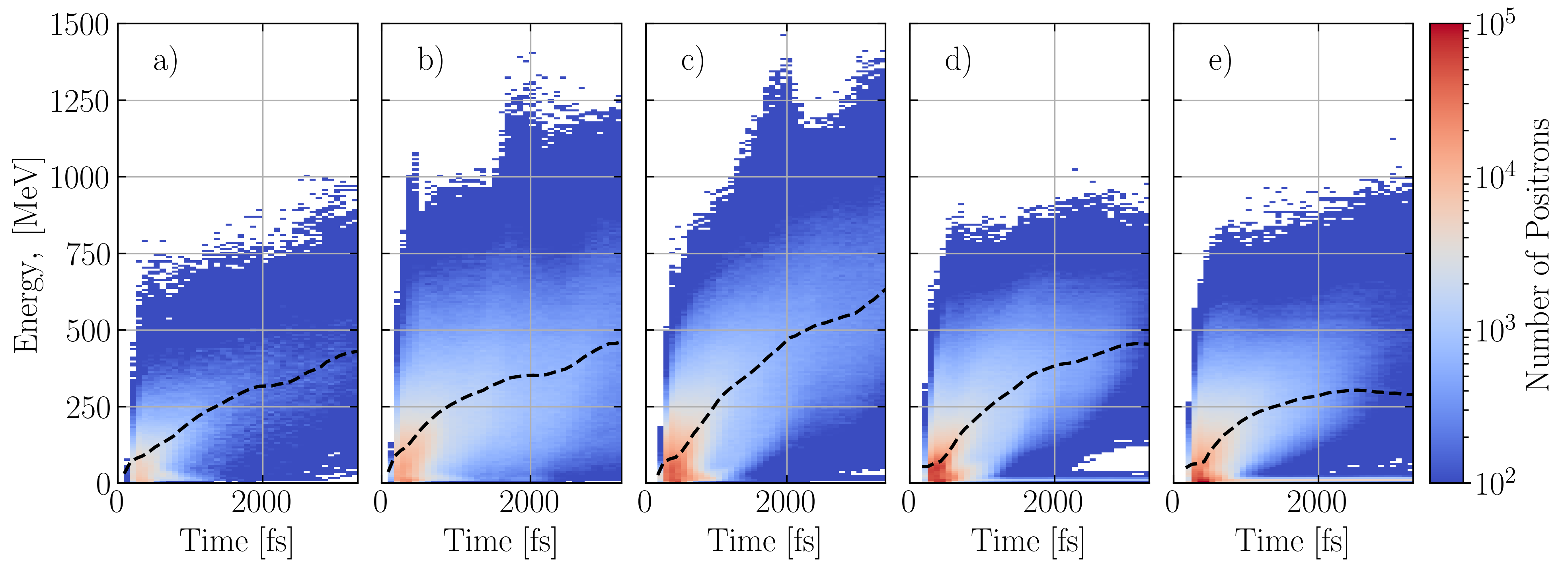}
\caption{Evolution of the positron distribution after creation and during its acceleration to a millimetre of propagation for the simulation with a foil at a) 400, b) 600, c) 800,d) 1000 and e) 1200 microns inside the plasma channel. A black dashed line is added representing the mean energy of the positron beam.}
\label{positron_beam_properties}
\end{figure}  

In Fig.~\ref{positron_beam_properties}, one can observe how the energy distribution of the positron beam evolves. As the positron beam propagates through the plasma, it gains energy via DLA. Besides, the positron beam charge decreases as it propagates (this can be noticed from frame a) to e)).  The energy gain is different for each case. In each case, the average energy gain for the positron beam is between $~300$ to $~600$~GeV/m. The simulation with the foil positioned at $600~\mu$m inside the channel shows a higher spread of the positron energy spectra. Notably, the simulation with the the foil positioned $800~\mu$m inside the channel has the largest energy gain which can be attributed to the electrons having achieved energy saturation before interaction with the target foil. It should be considered that the laser energy can be depleted quickly as it has already lost a considerable amount of energy after colliding with the foil.
\begin{figure}[ht]
\centering
\includegraphics[keepaspectratio,width = 0.9\textwidth,]{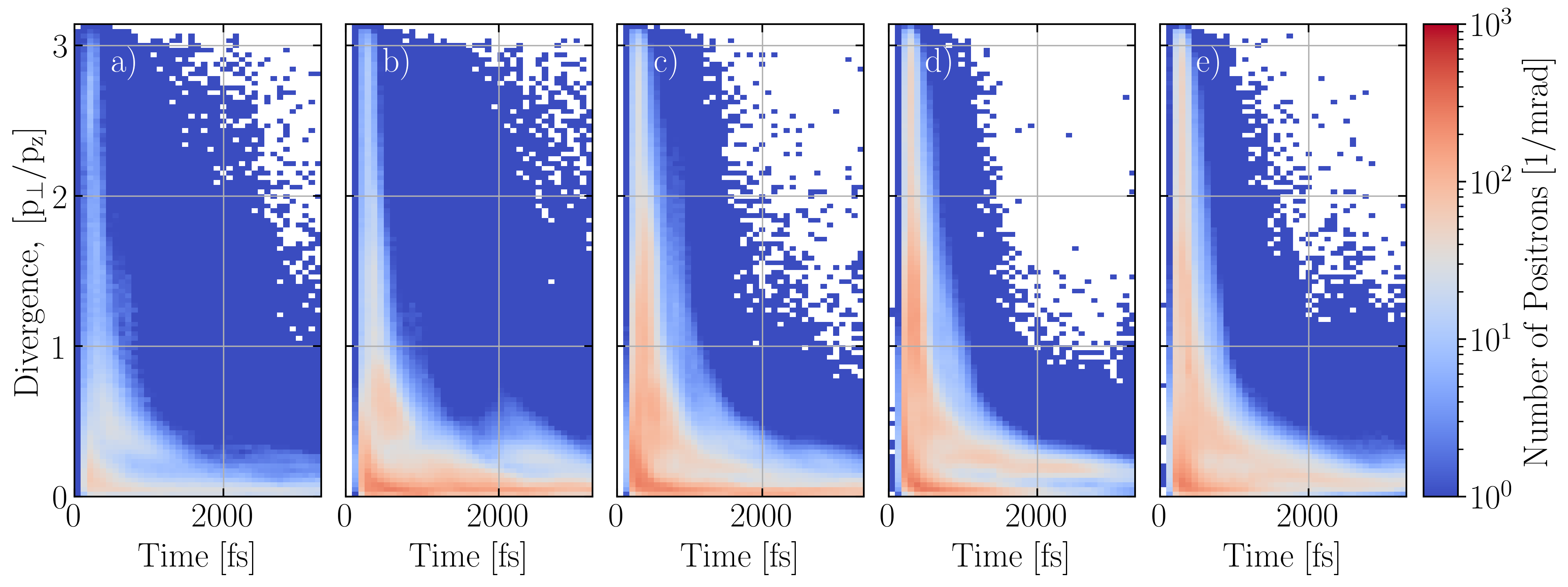}
\caption{Evolution of the positron beam divergence for the simulation with the foil located at a) 400, b) 600, c) 800, d) 1000 and e) 1200 $~\mu$m inside the plasma channel. We define $p_\perp = p_r = \sqrt{p^2_x + p_y^2}$}
\label{divergence_positrons}
\end{figure}

The divergence of the positron beam can also be studied. Figure~\ref{divergence_positrons} shows how the beam divergence evolves. When the positrons are created, they are directed in a vast range of angles due to the nonlinear interaction between the laser, the electron beam and the target, as seen in all frames. This high initial divergence leads to the loss of some positrons while the focusing fields successfully capture others. This could be attributed to some of the positrons with a higher divergence having overcome the focusing fields and other positrons gain higher longitudinal momenta, lowering their divergence. As a result, most positrons have a divergence of 120 millirad at the end of the simulations. The simulation with the foil position at $800~\mu$m has an average divergence of 85 millirad. Their divergence improves as they are accelerated in the longitudinal direction. This can be confirmed in Fig.~\ref{divergence_positrons}.

\subsection{Thicker targets}

We can potentially create more pairs by leveraging a more extensive interaction time with a target by increasing its thickness. In the previous sections, the aluminium foil thickness is $240~$nm. This section presents four simulations for aluminium foil thicknesses: $240, 500, 600$ and $750~$nm at $800~\mu$m inside the plasma channel. After the interaction with the aluminium foil, the simulations are set to conclude after a millimetre of laser propagation. Using  Eq.~10, we can expect the laser to have lost approximately $20~\%$ of its energy when it hits the target.

\begin{figure}[ht]
\centering
\includegraphics[keepaspectratio,width = 0.9\textwidth]{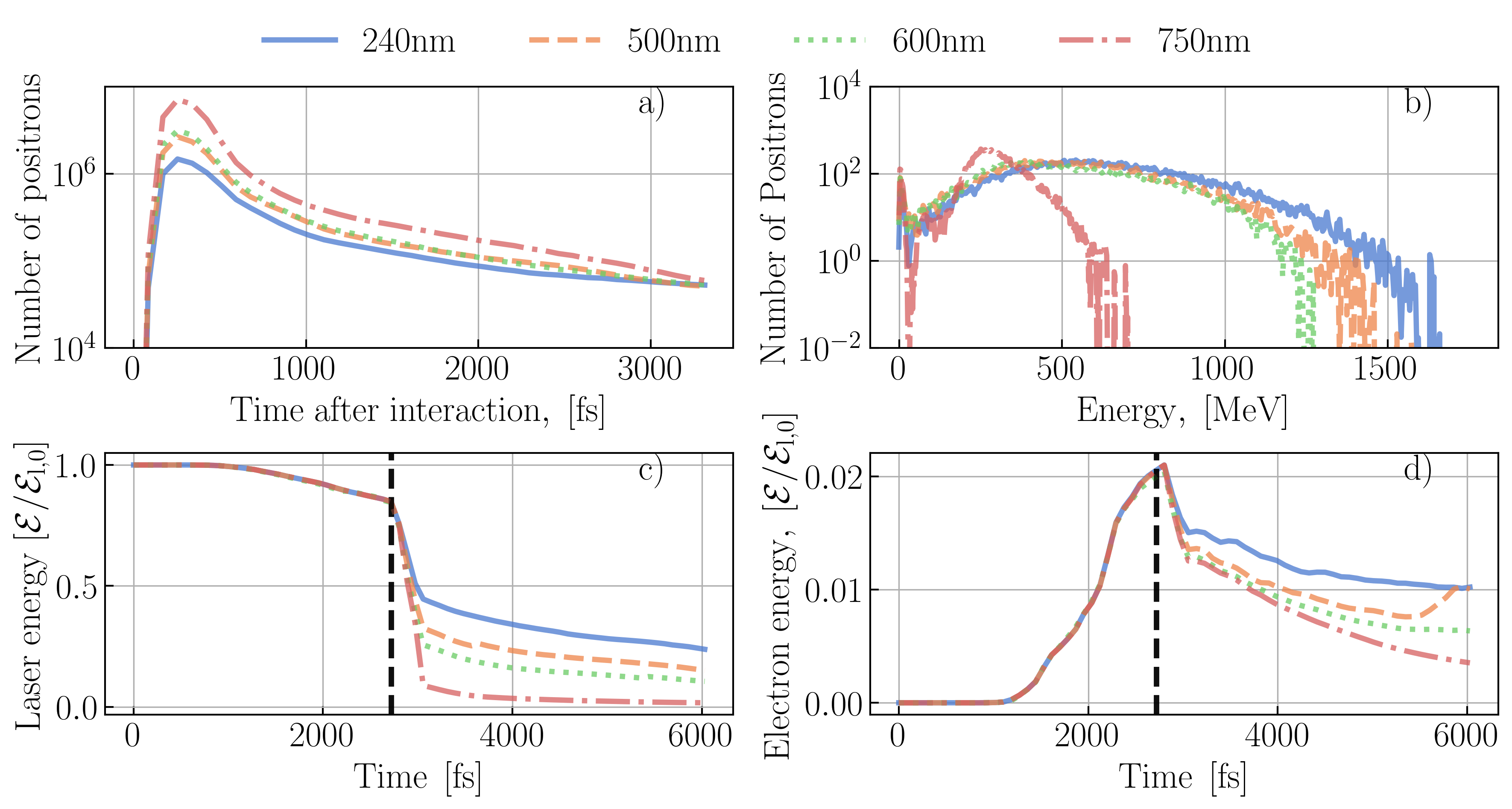}
\caption{Simulation results for simulations of varying aluminium target thickness from $240, 500,$ and $750$~nm. a) shows the number of positrons produced and retained. b) is the distribution of positrons as a function of energy at the last timestep of the simulation. c) is the laser energy revolution over time. d) is the evolution of electron energy over time. The black vertical dashed line signals the time at interaction with the foil (when the laser peak intensity hits the foil front).}
\label{varying_thickness}
\end{figure}

Figure~\ref{varying_thickness} displays results obtained with thicker targets. It can be concluded in panel a) that for a thicker target, more positrons are produced. For a target $240~$nm thick, we accelerate $1\times 10^{12}$ electrons,  we create approximately $1.5\times 10^6$ positrons; for a foil 3 times thicker, we produce over $7\times 10^6$ positrons. However, only 0.8$\%$ of the produced positrons are retained after a millimetre of propagation for the thickest target. Panel b) shows that the energy of the positrons, after a millimetre of propagation, is much lower than when thinner targets are used. This can be attributed to the laser nearly depleting after the collision and with the target, as confirmed in panel c). In addition, the energy of the electrons (consequently, the amount of electrons) dramatically decreases after interacting with the target, panel d). Depending on the thickness, the electrons lost $28\%, 35\%, 38\%$ and $41\%$ respectively when interacting with the target. In comparison, the laser loses $47\%, 61\%, 68\%$ and $89\%$ respectively. Depending on the positron-producing strategy, one could opt to obtain more energetic positrons by having a thinner target (between 240 and 500 nm). If one prefers more positrons, one should opt for a thicker target. However, it should be considered that the positrons will not be as energetic as in the thinner targets.

\section{Discussion}

This study expands upon previous work \cite{martinez_direct_2024} by presenting effective methods to optimize positron production and their subsequent acceleration and retention. In addition to the optimization study, we have introduced a simple analytical model which describes the energy transfer dynamics from the laser to the electrons. This model can be a valuable tool for further refining the optimization of positron production and acceleration techniques discussed here and in previous studies \cite{babjak_direct_2024}. It also adds to current DLA literature by providing insight into energy transfer from the laser to all the electrons.

The experimental realization of our proposed setup is feasible. As the ELI Beamlines L4 laser becomes operational, one could conduct this experiment using a plasma jet and a thin aluminium foil. Plasma densities from this work can be readily achieved with commercially available plasma jets \cite{sylla_development_2012}. Depending on the desired experiment, one can use different length plasma channels to either accelerate positrons to high energies or produce large quantities of positrons by varying the parameters, as highlighted in this paper. Naturally, one should choose the thin foil (in our case, aluminium), which is thin enough to allow for laser and beam transmission but sufficiently thick to facilitate positron production.

While we have focused on aluminium as the target material, it is worth noting that the setup is adaptable to higher-Z materials like copper or gold. These materials could potentially yield a quadratic increase in positron production. However, it is essential to consider that using higher-Z materials might influence the retention of the accelerated charge and the laser energy loss, aspects that warrant further investigation.

The primary objective of this paper was to provide a framework for optimizing positron production and retention by varying the key parameters: foil position, laser focusing position, channel wall density, and target thickness. While we did not explicitly explore parameters to maximize the energy of the accelerated positrons, it is essential to recognize that the acceleration mechanisms for positrons parallel those for electrons. Consequently, strategies outlined in previous literature to maximize electron energy \cite{babjak_direct_2024} can be readily adapted to enhance positron energy in our setup, after they have been created.

\section{Conclusion}

In this paper, we have demonstrated an effective method to improve the creation and acceleration of positrons in a single stage using DLA. We have shown that the injected charge can induce a charge inversion within the plasma channel, effectively trapping positrons. Moreover, we developed an engineering formula that illustrates the dynamics of energy depletion of the laser to the electrons. The rate at which the laser transfers energy to the electrons is proportional to the laser intensity and plasma density. Then, we showed that the focusing position of the laser within the plasma channel directly impacts the number of positrons created and retained. We also showed that positrons can originate both from the direct energy transfer from the laser into electrons at the front of the foil or by the DLA accelerated charge.
Furthermore,  we found that reducing the channel wall density enables the acceleration of a substantial electron charge (approximately 160~nC.), with a foil at $800$ microns inside the channel, producing approximately $1.5\times 10^{6}$ positrons. Our optimized setup with a foil placed $600$ microns inside a channel achieved an 8 times increase in positron retention compared to a previous study \cite{martinez_direct_2024}. We thereby show that the foil positioning is paramount. With further studies, we can potentially increase the amount of positrons created and retained even further. Placing a foil too close to the entrance of the channel prevents sufficient particle acceleration, whilst placing it too far away results in energy loss of the accelerated particles.
Finally, we describe the influence of the target thickness and show an increase in positron production at the cost of electron charge and laser energy. We have shown that we can create up to $7\times 10^6$ positrons for a thicker target, but the positron retention and energy gain decrease compared to thinner targets. There is an optimum value with the maximum amount of positrons created and the best retention rate at approximately $600$ microns inside a plasma channel for an aluminium target with a thickness of $240~$nm, which retains approximately $10\%$ of the produced pairs. This study presents a robust method of improving the production and retention of positrons in a single stage for experiments in upcoming Petawatt class lasers. This study can also be extended to calculate the optimization of maximum energies achievable for these positrons. 

\section{Acknowledgements}

The authors would like to acknowledge the support of the Portuguese Science Foundation (FCT) Grant No. CEECIND/01906/2018, DOI: PTDC/FIS-PLA/3800/2021 DOI: 10.54499/PTDC/FISPLA/3800/2021 and UI/BD/151560/2021 DOI: https://doi.org/10.54499/UI/BD/151560/2021. The supercomputing time was used by grant PARES (EuroHPC) and PILARS (RNCA).

\section{Bibliography}

\bibliographystyle{iopart-num}
\bibliography{main}

\end{document}